\pdfoutput=1
\documentclass[
	9pt,
	sigconf,
	screen,
	natbib=false,
	anonymous=false,
]{acmart}

\usepackage{verbatim} %
\usepackage{xspace}
\usepackage[binary-units]{siunitx}
\usepackage[USenglish]{babel}
\usepackage{amsmath}
\usepackage{mathtools}
\usepackage[font=footnotesize]{caption}
\usepackage[font=footnotesize]{subcaption}
\usepackage[inline]{enumitem}
\usepackage{float}
\usepackage{hyperref}
\usepackage{booktabs}		%

\usepackage{nth}			%
\usepackage{csquotes}		%

\usepackage[nolist]{acronym}

\usepackage[capitalise,nameinlink]{cleveref}
\crefname{section}{Sec.}{Sec.}
\Crefname{section}{Section}{Sections}
\crefname{equation}{eq.}{eq.}
\crefname{figure}{Fig.}{Fig.s}
\Crefname{figure}{Figure}{Figures}

\usepackage{float}
\usepackage{listings}

\newfloat{lstfloat}{htbp}{lop}
\floatname{lstfloat}{Listing}
\crefalias{lstfloat}{listing}

\usepackage[%
 backend=biber,
 datamodel=acmdatamodel,
 style=acmnumeric, %
]{biblatex}

\usepackage{tikz}
\usetikzlibrary{positioning, graphs, shapes.multipart, fit, calc, shapes.misc, matrix, shapes}
\usetikzlibrary{arrows, automata, plotmarks}

\newcommand*{\eg}{e.\,g.\@\xspace}
\newcommand*{\ie}{i.\,e.\@\xspace}
\newcommand{\R}{\mathbb{R}}
\newcommand{\N}{\mathbb{N}}

\newcommand{\node}{v}

\newcommand{\extractiondate}{August \nth{1} 2022}

\newcommand{\inlinecode}[1]{\texttt{#1}}

\begin{document}

\begin{acronym}[Derp]
\acro{p2p}[P2P]{peer-to-peer}
\acro{dht}[DHT]{distributed hash table}
\acro{ipfs}[IPFS]{Interplanetary Filesystem}
\acro{ipld}[IPLD]{Interplanetary Linked Data}
\acro{json}[JSON]{JavaScript Object Notation}
\acro{sfs}[SFS]{Self-Certifying Filesystem}
\acro{ipns}[IPNS]{Interplanetary Namesystem}
\acro{cid}[CID]{content identifier}
\acro{dag}[DAG]{directed acyclic graph}
\acro{protobuf}[Protobuf]{Protocl Buffers}
\acro{cdf}[CDF]{cumulative distribution function}
\acro{pdf}[PDF]{probability density function}
\acro{nft}[NFT]{non-fungible token}
\acro{url}[URL]{uniform resource locator}
\acro{uri}[URI]{uniform resource identifier}
\acro{ccn}[CCN]{content-centric networking}
\end{acronym}

\title[Dude, where's my NFT? Distributed Infrastructures for Digital Art]{Dude, where's my NFT?\\Distributed Infrastructures for Digital Art}

\acmYear{2022}\copyrightyear{2022}
\setcopyright{acmcopyright}
\acmConference[DICG '22]{3rd International Workshop on Distributed Infrastructure for the Common Good}{November 7, 2022}{Quebec, QC, Canada}
\acmBooktitle{3rd International Workshop on Distributed Infrastructure for the Common Good (DICG '22), November 7, 2022, Quebec, QC, Canada}
\acmPrice{15.00}
\acmDOI{10.1145/3565383.3566106}
\acmISBN{978-1-4503-9928-9/22/11}

\author{Leonhard Balduf}
\orcid{0000-0002-3519-7160}
\email{leonhard.balduf@tu-darmstadt.de}
\affiliation{%
	\institution{Technical University of Darmstadt}
	\city{Darmstadt}
	\country{Germany}}
\author{Martin Florian}
\orcid{0000-0003-2350-9283}
\email{martin.florian@hu-berlin.de}
\affiliation{%
        \institution{Weizenbaum Institute \& Humboldt-Universität zu Berlin}
	\city{Berlin}
	\country{Germany}}
\author{Björn Scheuermann}
\orcid{0000-0002-1133-1775}
\email{scheuermann@kom.tu-darmstadt.de}
\affiliation{%
	\institution{Technical University of Darmstadt}
	\city{Darmstadt}
	\country{Germany}}



\begin{abstract}
We explore issues relating to the storage of digital art,
based on an empirical investigation into the storage of audiovisual data referenced by \acp{nft}.
We identify current trends in \ac{nft} data storage and highlight problems with implemented solutions.
We particularly focus our investigation on the use of the \ac{ipfs},
which emerges as a popular and versatile distributed storage solution for \acp{nft}.
Based on the analysis of discovered data storage techniques,
we propose a set of best practices to ensure long-term storage survivability of \ac{nft} data.
While helpful for forming the \ac{nft} art market into a legitimate long-term environment for digital art,
our recommendations are also directly applicable for improving the availability and integrity of non-\ac{nft} digital art.
\end{abstract}

\maketitle

\acresetall

\section{Introduction}
\label{sec:intro}

Ensuring the availability and integrity of digital art is an inherent challenge~\cite{leeStateArtPractice2002},
as is its monetization.
\Acp{nft} promise to solve these issues.
This most modern form of digital art packaging attempts
to create digital scarcity and immutability by securing the \enquote{ownership} and metadata of digital art copies on a blockchain.
And it allegedly leverages distributed infrastructures for guaranteeing the long-term storage and availability of the actual content.
In this paper we focus on the availability and integrity challenges surrounding digital art,
as observable in the context of \acp{nft}.
We investigate how popular \acp{nft} \emph{actually} approach the storage question,
and to what extent the content they reference is \emph{actually} immutable.
Counter to reasonable consumer expectations,
we find that many of the most popular \acp{nft} store data on classical centralized platforms and permit changes to both metadata and actual content.
Still,
our empirical analysis of 1000 top-ranking \acp{nft}
also reveals multiple prominent \acp{nft} that successfully leverage on-chain storage and distributed storage infrastructures such as the \ac{ipfs}.
For content stored on \ac{ipfs},
we additionally take a look at \ac{ipfs} itself,
quantifying the extent to which stored \ac{nft} pieces are replicated (and hence, available) within the network.
Finally, we formulate a set of recommendations for \ac{nft} data storage.
The lessons we extract from the innovative approaches that (some) \ac{nft} projects take to digital art preservation are directly transferable to non-tokenized digital art as well.
Our key contributions are threefold:
\begin{enumerate*}
	\item a classification and empirical study of \ac{nft} data storage techniques
	\item an investigation of replication of popular content on \ac{ipfs}, and
	\item a set of recommendations for \acp{nft} data storage to enable long-term survivability
\end{enumerate*}
.

\section{NFT Storage in Current Practice}
\label{sec:nft_storage}

In the following,
we present an empirical study into how popular digital art \acp{nft}
approach the challenge of storing metadata and assets.
The choice of a storage method
has a determining impact on the long-term availability and (im)mutability properties of \acp{nft}
and the associated digital art.

\subsection{\Acp{nft}: Definition and Implementation}

\Acp{nft} are digital records that cryptographically connect a distinguishable digital token to an owner.
In practice, each individual token is identified by an integer referred to as token ID.
They are usually stored on a blockchain (the \enquote{chain}) and thus generally perceived as permanent.
This perception may extend to not only the digital token, but the assets referenced by it.
A highly prominent use case for \acp{nft} is to encapsulate ownership of \emph{digital art}
and audiovisual content more generally.
Other use cases include gaming (\eg, Mirandus\footnote{\url{https://mirandus.game/}}) and virtual reality --- or Metaverse --- (\eg, Decentraland\footnote{\url{https://decentraland.org/}}).

Most commonly, Ethereum~\cite{woodEthereumSecureDecentralised2014}
serves as the blockchain on top of which \acp{nft} are built\footnote{According to \url{https://dappradar.com/nft}}.
\ac{nft} implementations on Ethereum are typically governed by either
the \inlinecode{ERC721}~\cite{williamentrikenEIP721NonFungibleToken2018}
or \inlinecode{ERC1155}~\cite{witekradomskiEIP1155MultiToken2018}
smart contract interfaces.
These interfaces govern identification and ownership of \acp{nft}, including transfer thereof, as well as association of storage \acp{uri} to tokens, via optional metadata extensions.

\cref{lst:erc721} shows an excerpt of the commonly used \inlinecode{ERC721\-Metadata} extension interface.
The interface is used to attach a \ac{uri} for descriptive metadata to tokens.
The standard recommends that the metadata is a \acsu{json} document with a name, a description, and an image.
Commonly, the per-token metadata \ac{uri} is constructed on-the-fly from a base \ac{uri} and the numerical token ID, in order to save on-chain storage.
Audiovisual content, if any, is referenced via a \ac{uri} within the metadata document.
EIP721 provides no guidelines on how or where to store referenced content.
We investigate plausible techniques and the approaches that popular \acp{nft} projects take in the upcoming
\cref{sec:nft_storage}.

\begin{lstfloat}
	\vspace{-0.2cm}%
	\begin{lstlisting}
interface ERC721Metadata {
  ...
  function tokenURI(uint256 _tokenId)
  [...] returns (string);
}		
	\end{lstlisting}
	\vspace{-0.2cm}%
	\caption{The ERC721Metadata Interface.}
	\label{lst:erc721}
	\vspace{-0.2cm}%
\end{lstfloat}

The more recent \inlinecode{ERC1155} interface
allows tracking multiple distinct fungible and non-fungible tokens in one contract.
For \acp{nft}, this behaves exactly like \inlinecode{ERC721}
by using multiple tokens that each have a total supply of one.
An extension to assign metadata exists for \inlinecode{ERC1155} as well: \inlinecode{ERC1155Metadata\_URI}.
Here, the metadata \emph{must} be a \ac{json} document
with a predefined set of required keys.
An example is given in \cref{lst:eip1155_metadata}.

\begin{lstfloat}
	\vspace{-0.2cm}%
	\begin{lstlisting}
{
  "name": "Asset Name",
  "description": "Lorem ipsum...",
  "image": "http://example.com/{id}.png",
  "properties": { ... }
}
	\end{lstlisting}
	\vspace{-0.2cm}%
\caption{An example EIP1155 Metadata File.}
\label{lst:eip1155_metadata}
	\vspace{-0.2cm}%
\end{lstfloat}

It is noteworthy that a single contract, regardless of whether it implements \inlinecode{ERC721}, \inlinecode{ERC1155}, or both, can track multiple \emph{collections} of \acp{nft}.
The notion of grouping multiple tokens into a collection can exist on a contract-level, but is usually tracked on a higher level, \ie, on \ac{nft} marketplaces.

%
\subsection{Discovering Popular Digital Art \acp{nft}}
\label{sec:nft_storage:data_collection}

We obtained a list of the top 1000
most popular
\acp{nft}
on the OpenSea marketplace.
We used \enquote{last sale price}
as a proxy for popularity
and extracted the ranking from OpenSea on \extractiondate{}.
OpenSea is the most popular\footnote{by trading volume, according to DAppRadar, see \url{https://dappradar.com}} \ac{nft} marketplace and thus well-suited for our initial sample collection.

From our initial list of popular \acp{nft},
we proceeded to extract all \acp{nft} associated with digital art.
In a classification approach inspired by previous works such as \cite{nadiniMappingNFTRevolution2021},
we label each \ac{nft} as digital that art that has the primary purpose of referencing one or more audiovisual assets.
We explicitly exclude gaming and metaverse \acp{nft} from this class,
as well as utility tokens more generally.
%
%
%
%
%
%
%
$786$%
 \acp{nft} from our sample fit this classification%
\footnote{
  In addition, we classify
$179$%
 \acp{nft} as metaverse-related,
$19$%
 as game-related,
$8$%
 as collectibles,
  and %
$8$%
 as utility tokens.
}.
They are managed by a total of %
$51$%
 smart contracts,
many of which represent \ac{nft} collections with multiple representatives within our sample.
The subsequent discussion is based on these %
$786$%
 \acp{nft}.

\subsection{Metadata and Asset Storage}
\label{sec:nft_storage:where}

We proceed to classify digital art \acp{nft} in our sample based on
the storage system they use for storing metadata and assets.
We collected data stored in the \ac{nft} smart contracts,
\eg, data exposed through the EIP721 or EIP1155 standards,
and manually inspected custom contract implementations for data stored on-chain.

Motivated by the goal of digital art preservation,
we first identify what data is necessary to \emph{reconstruct} the artwork.
For example, generative art projects may choose to store source code on-chain, and provide a pre-rendered version of the artwork on cloud storage for convenience.
We classify this case as being stored on-chain, since reconstruction from on-chain data is possible.
It is potentially necessary to also preserve a runtime environment for this data,
but this falls outside the scope of this work.

We differentiate between three main storage approaches:
cloud storage, on-chain storage, and decentralized storage systems --- \ac{ipfs} and Arweave.
Their properties with regards to integrity and availability vary greatly:

\begin{itemize}
  \item \textbf{Cloud storage} offers no hard guarantees --- stored data can be mutated or (re)moved at any time.
  \item \textbf{On-chain storage} guarantees
    result directly from the security properties of
    the underlying blockchain system.
    Data availability and data integrity are among the core features
    of permissionless blockchain systems and it is a reasonable assumption that popular systems like Ethereum
    will remain intact --- and effective at providing availability and integrity for stored data --- for many years to come.
  \item \textbf{Decentralized storage systems} can have varying properties with regards to integrity and availability.
  The systems used by the digital art \acp{nft} in our sample rely
  on cryptographically-generated content identifiers:
  links into these storage systems correspond to cryptographic hashes of the stored content.
  In effect, changes to the original content can be detected as long as the integrity of the \ac{uri} of the asset is maintained.
  The availability of content stored in decentralized data storage systems
  is highly dependent on system particularities and the popularity of the stored content.
  In the upcoming \cref{sec:ipfs}, we report on a deeper investigation into how the \acp{nft} in our sample
  are inserted into the popular \ac{ipfs} network,
  and to what extent they are %
  replicated there.
\end{itemize}

The results of our classification are shown in \cref{tab:art_nft_storage_locations}.
To our surprise, the metadata and assets of more than half of our sampled digital art \acp{nft}
are already stored on-chain.
This is partly due to our manual inspection of contracts,
which reveals that various generative art projects store their source code on-chain in addition to providing pre-rendered versions of the artwork via other storage systems.
We discuss a representative example in \cref{sec:on_chain_storage}.

We attempted to resolve all metadata \acp{uri} pointing to cloud storage and decentralized storage systems on \extractiondate{}.
We discovered one case in which a metadata \ac{uri} pointing to a cloud endpoint could not be resolved (cloud-dead).
Keeping in mind that our sample is based on the top 1000 most valuable (based on our data)
\acp{nft} on OpenSea and that the availability of intact metadata is crucial to reconstructing the assets embodied by an \ac{nft}, this is a troubling result.
OpenSea continued to provide a cached versions of both metadata and assets for the \ac{nft} in question.
Relying on the caching mechanisms of \ac{nft} marketplace platforms for availability or integrity
is highly problematic, however.
\begin{table}
\centering
\caption{Metadata and Asset Storage Locations for Digital Art NFTs} 
\label{tab:art_nft_storage_locations}
\begin{tabular}{llr}
  \toprule
Metadata & Asset(s) & Count \\ 
  \midrule
on-chain & on-chain & 412 \\ 
  ipfs & ipfs & 254 \\ 
  cloud & cloud &  79 \\ 
  cloud & ipfs &  17 \\ 
  on-chain & cloud &  11 \\ 
  ipfs-dead &  &   7 \\ 
  arweave & arweave &   3 \\ 
  cloud & on-chain &   2 \\ 
  cloud-dead &  &   1 \\ 
   \bottomrule
\end{tabular}
\end{table}

\subsection{(Im)mutability}
\label{sec:nft_storage:immutability}

It is a reasonable assumption that a digital artwork, and especially a digital art \ac{nft},
should be immutable and maintain its original state.
In some cases, \acp{nft} do enable mutability, however.
While mutability might be a deliberate aspect of a digital artwork,
digital art that is meant to be immutable
should also be stored in an immutable manner. %

\subsubsection{Metadata mutability}%
\label{sec:nft_storage:immutability:metadata}

The most evident cases in which the immutability of assets referenced by an \ac{nft} is not secured
are \acp{nft} that use cloud storage for storing metadata or assets,
without any additional precautions such as recording a cryptographic hash of referenced data at a findable on-chain location
It must be noted, however, that it is not a \emph{sufficient} criterion for immutability
that the metadata and assets of an \ac{nft} are stored in an integrity-securing manner.
In many cases, an \ac{nft}'s smart contract might allow the metadata \ac{uri} itself to be modified,
thereby enabling a mutability of both metadata and assets.
Even though changes are visible on the blockchain, it is unreasonable to expect consumers to keep track of \ac{nft} contracts.

We examine \acp{nft} in our sample for transactions changing the metadata
\ac{uri} of tokens.
In general, neither EIP721 nor EIP1155 specify how \acp{uri} should be stored or updated, but we found that many contracts expose functions of the form \inlinecode{(set|update)[base](URI|uri)}.
By processing transaction data on the Ethereum blockchain,
we found that out of the 
contracts examined, $19$ changed the \ac{uri} of token metadata at least once.
Notably, we found one case where a base \ac{uri} was changed from a cloud provider to \ac{ipfs}, and later back to another cloud provider.
Closer examination of this case revealed that the metadata contained a
link
to the project homepage, which later changed location.
OpenSea still caches the \ac{ipfs} \ac{uri} for this project, highlighting the problem of cache inconsistency.
Additionally, we found one case where the base \ac{uri} of a project was temporarily set to an invalid \ac{uri}, which was corrected
around
$4$ hours later.

\subsubsection{Deliberate mutability}%
\label{sec:nft_storage:immutability:deliberate}

Having surveyed the state of mutability via \ac{uri}-changing transactions, we then explored whether \ac{nft} art projects are intentionally mutable.
For that, we examined each project and decided whether, from a consumer perspective and with reasonable time invested to research them, they presented themselves as mutable.
We conclude that %
$782$%
 of the digital art \acp{nft} in our sample are immutable, and only %
$4$%
 are deliberately and understandably not.
These projects communicate their mutability openly:
One of them is an evolving art piece, which will be updated by the artist over time.
Another one is the MoonCats\footnote{\url{https://mooncat.community/}} project, which allows owners to dress their virtual cats with accessories, which ultimately show up in the \ac{nft}-referenced image, visible on marketplaces and collector sites.
The latter is especially interesting, since this functionality is implemented via smart contracts, including all assets necessary to construct the images.
Artists can create accessories, which are (after a review) stored on-chain and made available to be bought and donned by owners of cats.
Internally, accessories are %
positioned and layered upon the base cat image to construct the resulting image.

Notably, a few of the cases that we perceived as immutable, but exhibited mutability through changing \acp{uri}, did so for marketing purposes: they changed the \ac{uri} from a preview or \enquote{teaser} version of the collection to the complete data on launch.
In most cases, however, mutability seems to be a byproduct of technical issues, such as the changing-homepage case laid out earlier.
%
%

%
%
%

%
\section{Focus: On-chain Storage}
\label{sec:on_chain_storage}

\Acp{nft} that feature on-chain storage of metadata or assets are typically based on a custom smart contract implementation.
Instead of generalizing, we will therefore focus on a specific example case:
the popular \ac{nft} collection \emph{CryptoPunks}.

CryptoPunks are a set of $10,000$ unique, $24 \times 24$ pixel character images.
There are, historically, multiple versions of the contract deployed on Ethereum\footnote{
\url{https://github.com/cryptopunksnotdead/punks.contracts}
}.
The original contract,
contains a SHA256 hash
of an image sprite containing all $10,000$ $24 \times 24$ pixel avatars, with transparent background.
The image itself is hosted in multiple locations off-chain, and the token ID of the \ac{nft} provides an index into the correct position in the image.

In August 2021, a contract to reconstruct the image data was deployed.
The contract %
builds images by compositing a \enquote{punk} from eight assets, which are each indexed through one byte.
The assets themselves are encoded compressed as a series of three-byte tuples, specifying coordinates and colors to apply.
Each tuple can paint an area of $2 \times 2$ pixels, which allows for assets of various sizes and positions to be encoded efficiently.
The contract stores a total of $133$ assets of various sizes, $100 \times 100 \times 8$ bytes of punk-data, a series of precomputed alpha-blended colors and a palette of base colors.
The developers spent $73$ million Gas on deploying the contract and data\footnote{\url{https://www.larvalabs.com/blog/2021-8-18-18-0/on-chain-cryptopunks}},
resulting in a total fee of approximately $4.05\;\text{ETH} \approx \text{USD}\,12\,205$ (at the time).

Overall, even though the contracts are not linked and thus some manual intervention is necessary, it is possible to retrieve images from on-chain data only.
This is impressive, and sets a good example for future projects.
It must be noted, of course, that the pixel-art style, composability from assets, and relatively small image size of $24 \times 24$ pixels are enablers for this feature.
Even so, the deployment incurred significant costs.
It still seems unlikely that large rasterized images will be stored on-chain in the future.

The CryptoPunks case exemplifies core characteristics of the on-chain storage approach for digital art \acp{nft}.
In summary:

\begin{itemize}
  \item The integrity and availability of embodied art pieces or codes is guaranteed as long as the underlying blockchain is intact and accessible.
  \item The involved smart contracts are non-standard,
    implying higher technical hurdles,
    respectively higher security risks.
  \item Due to the high cost of storing data on public blockchain networks like Ethereum,
    on-chain data storage is likely viable only for digital art with low storage space requirements,
    such as pixel art
    and generative art.
  \item Even though core data required for the reconstruction of the artwork is obtainable from the blockchain,
    auxiliary dependencies might still be required.
    We conclude that even on-chain storage should be paired with a holistic digital art preservation concept
    if long-term availability is a goal.
\end{itemize}

\section{Focus: Storage on IPFS}
\label{sec:ipfs}

In the following,
we examine data stored on \ac{ipfs} in greater detail.
\ac{ipfs} is the most popular distributed storage system for \acp{nft} at the moment, see \cref{tab:art_nft_storage_locations}.
We are furthermore able to leverage existing methods and tooling~\cite{henningsenMappingInterplanetaryFilesystem2020a,baldufMonitoringDataRequests2022}
for gathering detailed information about storage locations and data distribution within the \ac{ipfs} network.

\subsection{Background}

\Ac{ipfs} is a popular \ac{p2p} data storage and retrieval system~\cite{benetIPFSContentAddressed2014,baldufMonitoringDataRequests2022}.
It uses content addressing via so-called \acp{cid}.
A \ac{cid} is formed based on a cryptographic hash of the addressed data,
hence pointing to \emph{immutable} data.
Directories and large files are stored as \acp{dag} of hash-addressed blocks.

Downloading data is a two-step process: First, \emph{providers} are sought via a one-hop broadcast mechanism and a \ac{dht}-search~\cite{henningsenMappingInterplanetaryFilesystem2020a, baldufMonitoringDataRequests2022}.
Then, requests to download the data are sent to providers.
Downloaded content is, by default, cached locally and (re)provided upon request.
This effectively causes an increase in replication for data items that are frequently requested.

It is possible to add a whole directory of files to \ac{ipfs},
so that only the \ac{cid} of the directory (the root of the \ac{dag}) must be stored.
Individual files, \eg, metadata for individual tokens of a collection, can then be addressed relatively to this \ac{cid}.
Resolving such directories on \ac{ipfs} requires downloading the directory block, which contains \ac{cid} links to the files, and then downloading the file.
This is suboptimal for preservation since both blocks need to be downloadable.

Often, content is not requested via \ac{ipfs} directly but via a HTTP-\ac{ipfs} gateway, which fetches content via \ac{ipfs} and returns it via HTTP.
Public gateways are provided by the community.
This is convenient since it does not require running an \ac{ipfs} node locally, and public gateways are generally well-connected and quick to fetch content.
It does, however, create dependence on centralized infrastructure \cite{baldufMonitoringDataRequests2022}.

Since content is downloaded directly from nodes on the network, it is available for as long as it is held
by at least one nodes that is online and discoverable.
Some companies %
provide paid \emph{pinning} services, which store content on their nodes, making it available to the \ac{ipfs} network.

\subsection{Empirical Study}

We focus on the %
$254$%
digital art \acp{nft} in our sample that use \ac{ipfs} for storing both metadata and assets.
Out of those, one contract did not conform to the metadata specification and returned an image instead
--- we ignore the associated \acp{nft} in the following.

\subsubsection{\ac{cid} Encoding}

In principle, a \ac{cid} addressing raw binary data can be stored in 32 bytes,
by encoding only the hash
and generating a valid (textual) \ac{cid} from it on the fly.
Still,
we found that \emph{all} of the %
$254$%
 investigated \acp{nft} used regular text-coded \ac{ipfs} \acp{uri}
requiring upwards of 50 bytes of on-chain storage. %
Many contracts additionally use a hard-coded \ac{url} to a public gateway server,
in the form \inlinecode{https://<gateway>/ipfs/<cid>},
instead of a regular \inlinecode{ipfs://<cid>} \ac{uri}.
While this makes it more convenient for users to access the referenced metadata and assets,
it is wasteful in terms of on-chain storage.
Additionally, when users resolve \ac{ipfs} \acp{cid} exclusively via public HTTP gateways,
the expectable levels of availability and integrity degrade to the level of cloud-base storage~\cite{baldufMonitoringDataRequests2022}.

\subsubsection{Data Availability}%

We extract a list of all \ac{ipfs} metadata \ac{uri} from all

considered \ac{nft} smart contracts,
which we resolve to obtain a list of all metadata and asset \acp{cid} for the
$254$%
 digital art \acp{nft} in our sample.
We then conducted a measurement study to determine the availability and replication of these data items.
Specifically, we attempt to identify all providers for the referenced data items at six-hour intervals over the course of one week.
We do this through searching the \ac{dht}, using a well-connected, non-NATed node located in Germany.

Notably, the \ac{dht} is not the only discovery mechanism in \ac{ipfs} --- there is also a one-hop broadcast mechanism via the Bitswap protocol~\cite{baldufMonitoringDataRequests2022}.
We plan to extend the study presented here with Bitswap discovery results in the future.

\begin{figure}[ht]
	\centering%
	\vspace{-0.2cm}%
	\input{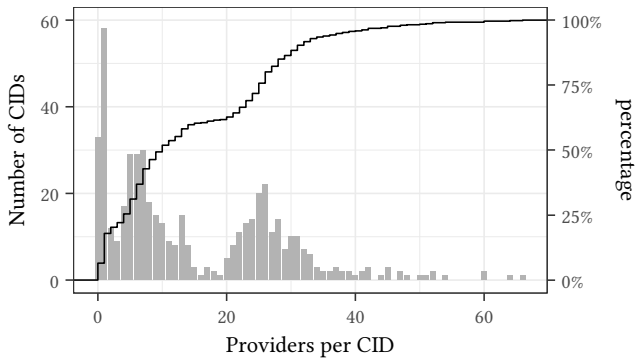}\unskip%
	\vspace{-0.3cm}%
	\caption[Replication per CID]{
		Number of Replicas per CID.
		Based on one week of DHT crawls every six hours, counting unique (\ac{cid}, peer ID) tuples.
	}%
	\label{fig:block_availability_by_cid}
\end{figure}

We consider unique (peer ID, \ac{cid}) tuples over all measurements and investigate how often a \ac{cid} appears within the set of all tuples, \ie, how many times a \ac{cid} is replicated.
The results, visualized as a histogram of the discrete counts as well as an ECDF, are shown in \cref{fig:block_availability_by_cid}.
Among other tings, it can be seen that
50{\%} of \acp{cid} were provided by less than ten nodes each.
(Over the course of the whole week, not just at one instant!)
Additionally, for a surprising $33$ \acp{cid} no providers could be discovered via the \ac{dht} at all.

These results are to some extent puzzling ---
\ac{nft} platforms such as OpenSea do not typically communicate to users (in our experience)
that some \ac{nft} might be unavailable beyond the current platform's cache.
It is possible that our results hint at problems on \ac{ipfs}'s \ac{dht} layer,
i.e., that some providers for a given \ac{dht} can only be discovered via one-hop broadcasting.
Our results from \cref{sec:nft_storage}, in which we were unable to resolve only 7 \acp{cid}
(for a different time frame and not restricting the content search to the \ac{ipfs} \ac{dht}),
also point to this conclusion.
Further studies are necessary,
involving both \ac{dht} lookups and broadcast-based searches from nodes with a large number of simultaneous peers
(e.g., modified nodes as in \cite{baldufMonitoringDataRequests2022}).

%

\section{Related Work}
\label{sec:related}

\textcite{nadiniMappingNFTRevolution2021} investigate \ac{nft} trades from 2017 to 2021 on Ethereum and WAX.
They describe trade networks and investigate visual similarity between items as indicators for price.
A classification of \acp{nft} by use case
(\eg, profile pictures)
is performed,
however without a consideration of storage and immutability questions.
\textcite{vasanQuantifyingNFTdrivenNetworks2022} explore \acp{nft} from a network-scientific view.
They note that, due to the inherent transparency of the underlying blockchains, novel insight into the art ecosystem is possible.
The authors perform extensive analyses of trader-, collector-, and artist networks on the \emph{Foundation} platform.
Storage and immutability challenges are, again, not explicitly addressed.

\textcite{dasUnderstandingSecurityIssues2022} showcase a wide range of security issues in the \ac{nft} ecosystem.
They analyze popular \ac{nft} marketplaces and identify various security issues pertaining to, \eg, verification of authenticity, preventing fraud, etc. 
Most relevantly to our work,
they showcase the persistence of off-chain NFT data as an unsolved issue.
They analyze the storage location and availability of NFT metadata and assets based on a
dataset of 12,215,650 \acp{nft} (of all types) from OpenSea.
They find that only a fraction of the referenced content is stored on \ac{ipfs}, and that a large portion of data stored on other systems is inaccessible already.
Interestingly,
while the research question behind this part of their work is similar to ours,
their results and conclusions differ from what we outline in \cref{sec:nft_storage}.
This fact can be explained by differences in the dataset used in both studies
--- we focus only on especially popular \acp{nft}, for example ---
as well as our detailed investigation of contract implementations to identify on-chain storage solutions.
We extend the work of \citeauthor{dasUnderstandingSecurityIssues2022} by looking at \ac{nft} storage approaches and their impact in greater detail.
We furthermore contribute an in-depth exploration of \ac{ipfs} as a storage medium for \ac{nft},
including an empirical study on the \ac{ipfs} network itself that aims to determine the level of replication of surveyed \acp{nft}.
Empirical studies on \ac{ipfs} have been performed in the past
\cite{henningsenMappingInterplanetaryFilesystem2020a,baldufMonitoringDataRequests2022}
and offer a valuable methodological foundation for our work.
To the best of our knowledge,
however,
no published works exist that evaluate the level of replication of a specific set of objectively popular
content within the \ac{ipfs} network.

Our work is tangentially related to the study of digital art preservation
which arose with the introduction and use of digital systems in the creation and storage of art in the second half of the \nth{20} century.
\textcite{leeStateArtPractice2002} give an overview of techniques used for digital preservation.
Since both \acp{nft} and \ac{ipfs} are young technologies,
they might not yet be in the focus of digital art preservation efforts.
With this work,
we hope to give guidelines on best practices for artists and creators to make digital preservation efforts easier in the future
--- both of digital art \ac{nft} but also of non-tokenized art projects that want to leverage \ac{nft}-related technologies for improved availability and integrity.

\section{Conclusions and Outlook} %
\label{sec:conclusion}

In this work, we surveyed the current state of \ac{nft} data storage empirically,
showcased examples and concrete implementations of various storage approaches,
and measured replication of data on \ac{ipfs}.
Based on the preceding discussion and our empirical results,
we can arrive at a list of recommendations for upcoming digital art \acp{nft}:

\begin{itemize}
	\item On-chain storage should be the gold standard, and considered for any data small enough to allow for this.
	Metadata files can already be stored on-chain using \inlinecode{data} \acp{uri}~\cite{masinterDataURLScheme1998}.
	This could further be optimizing by storing only relevant field values on-chain and constructing the metadata \ac{json} on-the-fly.
	On-chain storage of assets is only feasible for a subset of digital art, such as pixel art, ASCII art, or generative art.
	For the latter, dependency management and reproducibility should be considered, \eg through version-locking dependencies on-chain.
	
	\item Assets should be immutable by default, which realistically corresponds to user expectation.
	Mutability should be communicated explicitly.
	Mutations to metadata should be done transparently in a way that allows reconstruction from on-chain data.
	
	\item If on-chain storage is not an option, a content-addressed distributed storage system should be considered.
	For any off-chain storage solution, a hash of the referenced data should be stored on-chain.
	This hash can generally be stored binary and reconstructed on-the-fly.
	Distributed storage systems should be employed in a way that incentivizes replication of the data.
	Some storage systems explicitly promise long-term availability of data, which should be a goal.
\end{itemize}

Potential follow-up works include extending our study on the replication and availability of \ac{nft}-related content on \ac{ipfs},
to also consider content discovery mechanisms beyond \ac{dht} lookups.
Our insights on \ac{nft} storage practices might furthermore inspire the development of improved long-term storage approaches.
For example,
it is conceivable to devise a system that stores content on multiple storage systems in parallel.
Stored hashes might be used for automatically generating, or in some way authenticating,
content identifiers for multiple distributed storage systems
(\ac{ipfs} and BitTorrent, for example).
Improved storage approaches like the sketched idea
would benefit all forms of digital preservation efforts,
including the preservation of non-\ac{nft} digital art.

\begin{acks}
The authors would like to thank the anonymous referees for their valuable comments and helpful suggestions.
This work was supported by \grantsponsor{PL}{Protocol Labs}{https://research.protocol.ai/}
under Grant No.~\grantnum{PL}{PL-RGP1-2021-054}. %
This work was also supported by the \grantsponsor{BMBF}{German Federal Ministry of Education and Research (BMBF)}{https://www.bmbf.de/}
through funding for the Weizenbaum Institute for the Networked Society.
\end{acks}

\printbibliography

\end{document}